\documentclass[aps,twocolumn,showpacs]{revtex4}
\usepackage{graphicx}
\usepackage{dcolumn}
\usepackage{bm}
\usepackage{CJK}
\usepackage{amsfonts}
\usepackage{psfrag}
\usepackage{wrapfig}
\usepackage{subfigure}
\usepackage{makeidx}
\usepackage{multirow}
\usepackage{epsf}
\usepackage{hyperref}
\usepackage{amsmath}
\usepackage{cases}

\begin{document}
\title{Prolific interaction structures and shape-changing collisions in an erbium-doped fiber system}
\author{Liang Duan$^1$}
\author{Zhan-Ying Yang$^1$}\email{zyyang@nwu.edu.cn}
\author{Chong Liu$^{1}$}
\author{Wen-Li Yang$^2$}
\address{$^1$School of Physics, Northwest University, Xi'an
710069, China}
\address{$^2$Institute of Modern Physics, Northwest University, Xi'an
710069, China}

\date{November 27, 2015}
\begin{abstract}
Prolific interactions of nonlinear waves on a plane-wave background in an erbium-doped fiber system
are unveiled, based on explicit coexistence conditions extracting from the general higher-order solution of a coupled nonlinear Schr\"{o}dinger and the Maxwell-Bloch equations. In contrast to previous results, it shows that rich different types of nonlinear waves can coexist and interact with each other.
In particular, we reveal an interesting kind of shape-changing collision,
in which the multi-peak solitons exhibit intensity redistribution characteristics. This interaction involves the mutual collision between multi-peak solitons as well as the interplay of multi-peak soliton and other types of localized nonlinear waves (breather and antidark soliton).
We find that the shape-changing collision preserves the power of each multi-peak soliton, which is verified via optimal numerical integration for the energy of light pulse against the plane-wave background.
Our results demonstrate that when the interactions occur, each multi-peak soliton exhibits the coexistence of the shape change and energy conservation.

\end{abstract}

\maketitle

\section{Introduction}
Nonlinear wave interactions are an integral part of nonlinear wave
theory, which often exhibit highly nontrivial features \cite{NLW1,NLW2,NLW3,NLW4}.
The intraspecific interactions, including the well-known soliton interaction \cite{s1,s2,s3} and breather interaction \cite{b1,b2,b3,b4,b5,b6,b7}
with characteristic particle collision properties,
commonly appearing in scalar nonlinear wave evolution equations,
has been extensively studied in many fields both theoretically and experimentally. Another interesting case
is the interspecific interaction, i.e., interaction
between different types of nonlinear waves \cite{r1,r2,r3,r4,r5,r6,r7}. In fact, research on interspecies interactions is in an emerging state.
Recently, interactions between localized nonlinear waves (soliton, rogue wave, and breather)
and periodic waves (standard cnoidal waves) have been demonstrated in the standard scalar nonlinear Schr\"{o}dinger (NLS) systems \cite{r5,r6,r7}.
Essentially, these interactions are shape-unchanging.

In contrast to the standard scalar NLS case, interactions, especially interspecific interactions exhibit structural diversity in
coupled multicomponent (CM) systems. It is because that CM systems possess some additional coupling parameters and
allow for interaction between different components, which potentially yield rich and significant interaction structures.
One of the most well-known examples is the shape-changing collision for the standard solitons in the coupled NLS systems \cite{t1,t2,t3}, revealed
by the analysis of intensity redistribution among the components.
More remarkably, rich interspecific interactions have been recently identified in the coupled NLS systems \cite{n1,n2,n3,n4,n5,n6,n7,n8,n9,n10},
such as attractive interactions between solitons and rogue waves \cite{n1,n2,n3,n4,n5,n6}, inelastic collisions between breathers and other-type waves
(solitons and rogue waves) \cite{n7}, and others \cite{n8,n9,n10}.
As a result, CM system opens a venue into rich interaction dynamics of nonlinear waves, which needs more exploration.

In the present paper we extend nonlinear wave interactions to a coupled nonlinear Schr\"{o}dinger and the Maxwell-Bloch (NLS-MB) model,
which describes optical pulse propagation in a resonant erbium-doped fiber \cite{m1,m2,m3}.
Recently, the well-known localized nonlinear waves, i.e., soliton, breather and rogue wave and their interactions with shape-unchanging structures were studied in \cite{p1,p2,p3,p4,p5}; different types of localized and periodic waves on a plane-wave background were also revealed \cite{chong}. The next step of interest and significance is to investigate whether these nonlinear modes can coexist and interact with each other. However, to our knowledge, the important interplays of these nonlinear modes have not been studied, not even for the possibility of the coexistence and interaction, but also property of interactions has not been analyzed at all.

Our treatment below goes as follows. In Section \uppercase \expandafter {\romannumeral 2},
exact nonlinear wave solutions on a plane-wave background are constructed; in particular, possibilities of coexistence and interaction between different-type nonlinear waves are analyzed in detail, based on the explicit coexistence conditions.
In Section \uppercase \expandafter {\romannumeral 3}, some striking coexistence and interaction, including the interspecies interactions between breathers and other types of nonlinear waves (periodic wave, antidark soliton, and multi-peak soliton), the mutual collision between multi-peak solitons, and the interplay of multi-peak soliton and antidark soliton, are presented. Especially, a kind of shape-changing collision, which is identified as the result of the self-shaping effects of multi-peak solitons, is revealed analytically and numerically. Moreover, it is demonstrated that the shape-changing interactions have no analogues in the standard scalar NLS systems. The final section presents our conclusions.

\section{nonlinear waves in NLS-MB system and interaction analysis}
We consider a resonant erbium-droped fiber system governed by a coupled system of the NLS-MB equations \cite{m1,m2,m3}
\begin{equation}
\begin{split}
&E_{z}=i\left(\frac{1}{2}E_{tt}+|E|^{2}E\right)+2P,\\
&P_{t}=2i\omega P+2E\eta,\\
&\eta_{t}=-(EP^{*}+PE^{*}),
\end{split}
\end{equation}
where $E(z,t)$ is the slowly varying envelope field; $P(z,t)$ is the measure of the polarization of
the resonant medium, which is defined by $P=v_1 v_2^*$; $\eta(z, t)$ denotes the extent of the population inversion, which is given
by $\eta=|v_1|^2-|v_2|^2$, $v_1$ and $v_2$ are the wave functions of the two energy levels of the resonant atoms;
$\omega$ is the carrier frequency, and the index $*$ denotes complex conjugate.

To study the prolific interaction structures between different types of nonlinear waves in an erbium-doped fiber system, we shall construct the exact general higher-order nonlinear wave solutions on a plane-wave background. Explicit exact first-order and second-order solutions are given in the Appendix.

The expressions of the solutions depend on the background wave amplitudes $a$, $k$,
the background wave frequency $q$, the carrier frequency $\omega$, and real parameters $a_j$, $b_j$ ($j=1,2$).
When $a_1=0$, the first-order solution will reduce to the case in Ref. \cite{chong}; different abundant types of nonlinear structures including multi-peak soliton, periodic wave, antidark soliton, and W-shaped soliton (as well as the known bright soliton, breather, and rogue wave) were revealed by the analysis of velocity
difference between hyperbolic and trigonometric functions.
However, the obvious drawback of this approach is that it does not allow us to investigate some new specific interspecies interplays.
For instance, we cannot obtain the interaction between multi-peak solitons and antidark solitons by nonlinear superposition of the first-order solutions with one selected value of $q$.

We note that the additional degrees of freedom provided by $a_1$, $a_2$ allow for more abundant interplays of
different nonlinear waves without affecting the wave types, which will be displayed in the following.
In this case,
we present existence conditions of nonlinear waves with a general and concise form, which is shown in Table \uppercase \expandafter {\romannumeral 1}.
The interesting finding is that the existence conditions of types of nonlinear waves are the same for an arbitrary-order solution.

Let us then reveal the potential possibility of the interplays of nonlinear waves in the system.
We find that not all of these nonlinear waves can interact with each other. Three interspecific interactions
(periodic wave, anti-dark soliton, and W-shaped soliton) do not exist in the system due to their
existence conditions. Additionally, without losing generality,
we omit the cases for rogue waves and W-shaped solitons, since they are merely the limiting cases of breathers and periodic waves, respectively.
In the remaining cases, our interest is mainly confined to providing some intriguing interaction patterns of the rich different-type nonlinear structures in the system,
which have not been reported before. Specifically, we present two types of interaction structures, including shape-unchanging and shape-changing interactions.
Remarkably, we find that the multi-peak solitons exhibit shape-changing properties
before and after interactions.

It is noteworthy that, if $k=0$, implying $P=0$, $\eta=0$, the NLS-MB model reduces to the standard NLS equation. In this case the multi-peak soliton, antidark soliton, W-shaped soliton, and periodic wave reported in this paper cannot exist in the standard NLS system, since the existence condition $k\neq0$ [see Table \uppercase \expandafter {\romannumeral 1}]. As a result, the interactions, which these waves are involved, are specific to the NLS-MB system, and could exhibit unique and significant dynamics properties.

\begin{table}[!hbp]
 \label{table1}
 \begin{tabular}{|c|c|}
 \hline
  Nonlinear waves type  & Existence condition \\
  \hline
  Breather and rogue wave & $b_j^2+A_j^2\neq k$~ ($A_j=a_j+\omega$)\\
  \hline
  Multi-peak soliton & $b_j^2+A_j^2=k$, $q\neq-2a_j$ \\
  \hline
  Periodic wave &  $b_j^2+A^2_j=k$, $q=-2a_j$, $a^2>b_j^2$ \\
  \hline
  Antidark soliton & $b_j^2+A_j^2=k$, $q=-2a_j$, $a^2<b_j^2$ \\
  \hline
  W-shaped soliton & $b_j^2+A_j^2=k$, $q=-2a_j$, $a^2=b_j^2$ \\
  \hline
\end{tabular}
\caption{Types of nonlinear waves in NLS-MB system with corresponding explicit condition.}
\end{table}
\begin{figure}[htb]
\centering
\subfigure[]{\includegraphics[height=40mm,width=48mm]{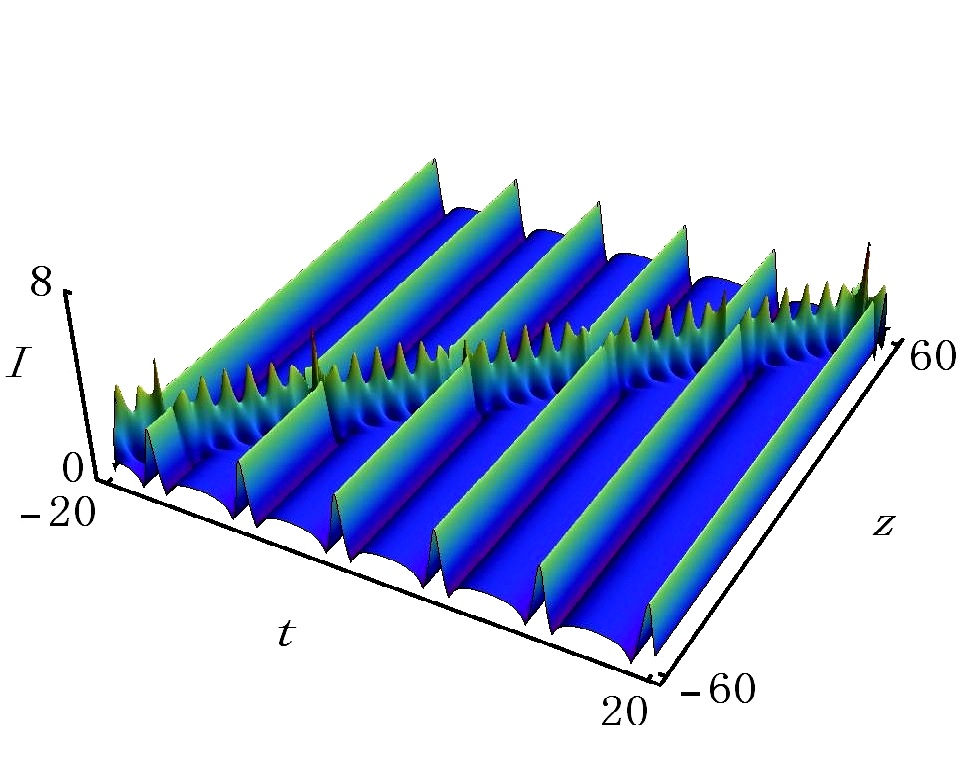}}
\subfigure[]{\includegraphics[height=40mm,width=36mm]{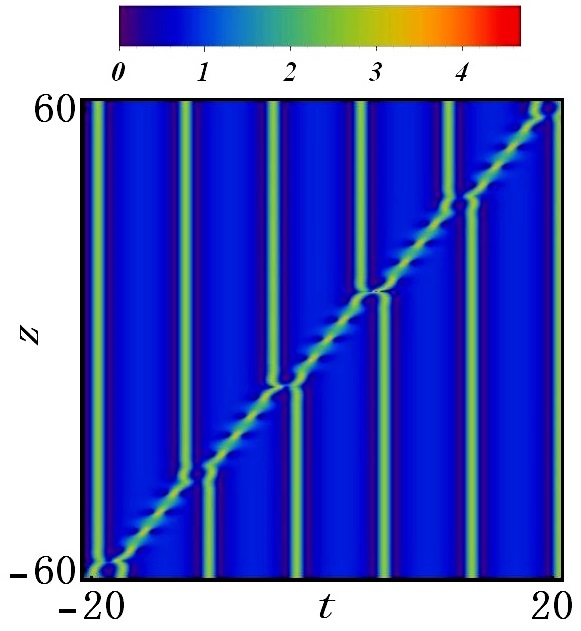}}
\subfigure[]{\includegraphics[height=40mm,width=48mm]{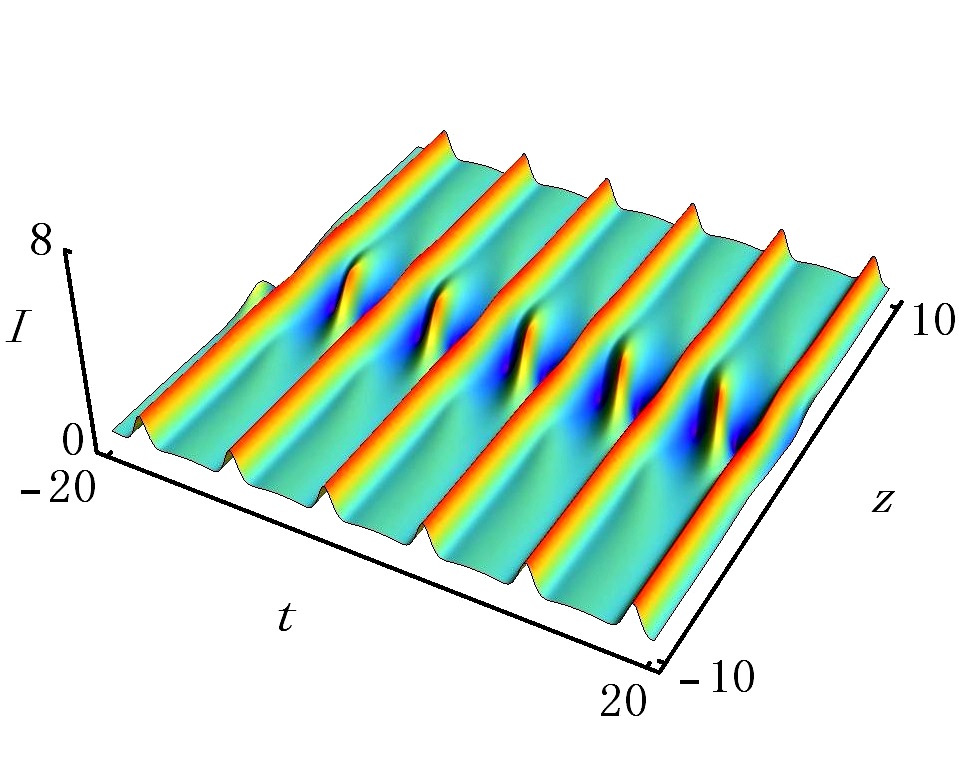}}
\subfigure[]{\includegraphics[height=40mm,width=36mm]{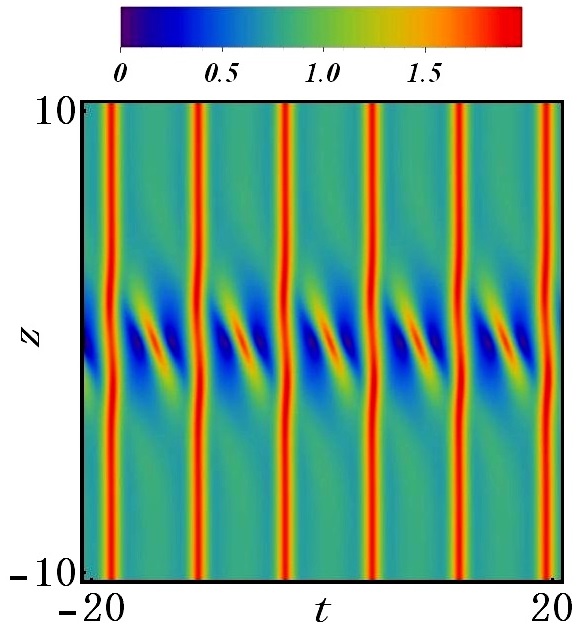}}
\caption{Shape-unchanging interactions between breathers and periodic waves,
(a) the Kuznetsov-Ma breather and a stationery periodic wave with $b_2=1.5$;
; (b) the Akhmediev breather and a stationery periodic wave with $b_2=0.5$. Others are $a=1$, $b_1=0.9$, $a_1=a_2=-1/3$, $\omega=1$, and $q=-2a_1$.}
\end{figure}

\section{characteristics of interplays}

\subsection{Shape-unchanging interactions between different types of nonlinear waves}
We first consider the interplays of breathers and other-type nonlinear waves.
We present some interesting coexistence and interaction structures, including the interaction between
breathers and periodic waves and the collision between breathers and anti-dark solitons. The corresponding typical
amplitude [$I(z,t)=\sqrt{|E(z,t)|^2}$] distributions are depicted well in Figs. 1 and 2.

Figure 1 exhibits the coexistences and interactions between breathers and periodic waves.
The corresponding coexistence conditions are chosen as $b_1^2+A_1^2=k, q=-2a_1$, $a^2>b_1^2$, and $b_2^2+A_2^2\neq k$,
based on the exact classification in Table \uppercase \expandafter {\romannumeral 1}. To better understand the
coexistences and interactions, we set the periodic wave stationary and select two well-know breathers, i.e.,
the Akhmediev breather (AB) \cite{AB} and the Kuznetsov-Ma breather (KMB) \cite{KMB} propagating on the periodic wave background.

Figure 1(a) shows,
the KMB propagation on a W-shaped wave train background. One should note that the nonzero velocity of the KMB is induced by
the structural parameter $\omega$, which is different from the case of the standard NLS system. Specifically, when the
KMB passes through the W-shaped wave train, it collides with each W-shaped wave and some large amplitude peaks are formed. The KMB and the W-shaped wave restore
the original shapes after each collision with a small phase shift. This indicates that the whole interaction structure is elastic.
Figure 1(b) illustrates the interaction between the AB and a periodic wave which are well coexisting.
The each peak of the AB is localized between two adjacent W-shaped waves. The periodic wave exhibits a slight snakelike oscillation
as it coexists and interacts with the AB, while the AB remains its innate structure.

For the choice of parameters $a^2>b_1^2$, and the rest remains the same, one obtains the interaction between breathers and antidark solitons.
Figure 2 depicts the corresponding interaction structure.
Although the waves are two distinct types of localized waves, the collision between them exhibits an elastic feature on the same plane-wave background.
From $z=-10$, two waves approach each other, at $z =0$ they collide with
each other and form a large amplitude peak. They then separate after the collision, to restore the original shape; there is no energy exchange during the collision.

\begin{figure}[htb]
\centering
\subfigure[]{\includegraphics[height=40mm,width=48mm]{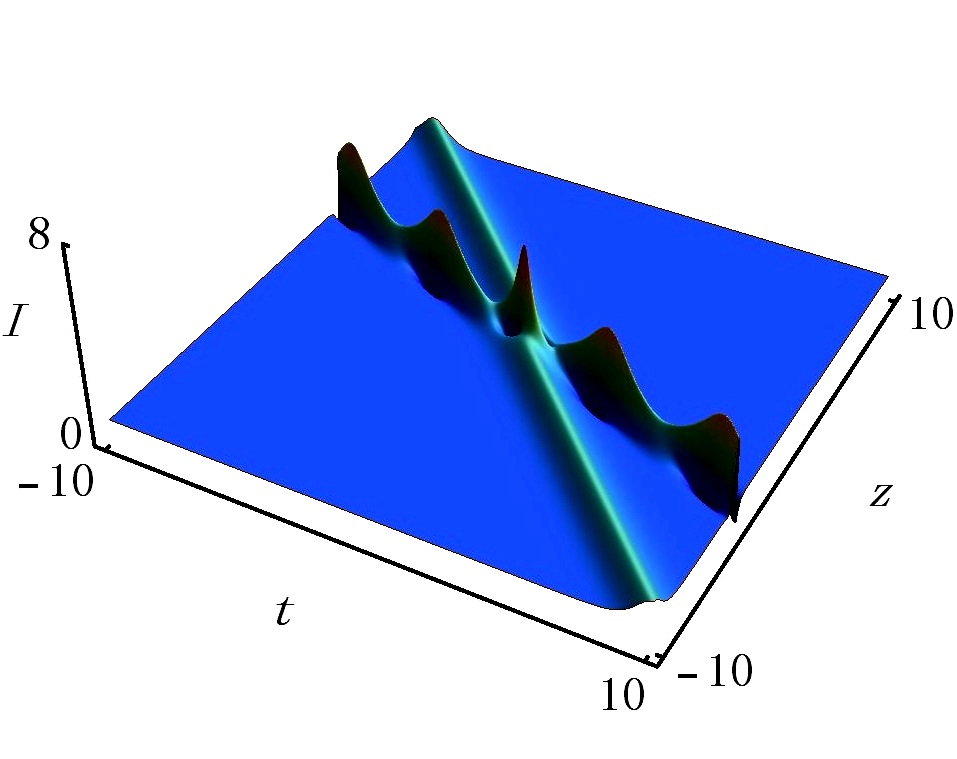}}
\subfigure[]{\includegraphics[height=40mm,width=36mm]{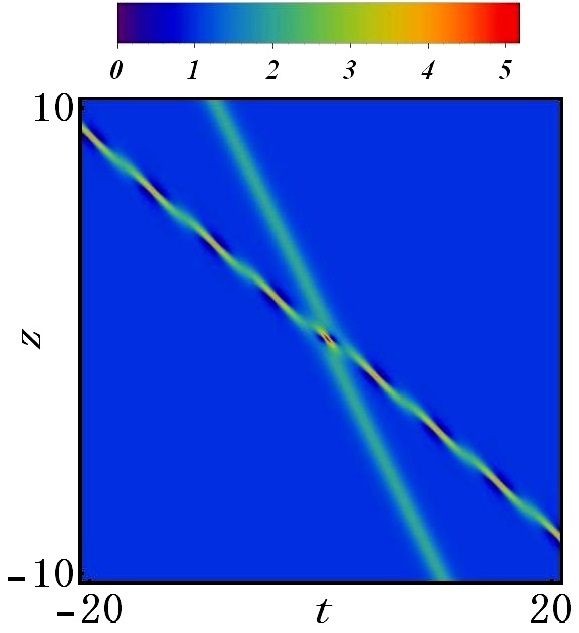}}
\caption{Shape-unchanging interaction between breathers and anti-dark solitons. The setup is $a=1$, $a_1=0$, $b_1=1.5$, $a_2=0.8$, $b_2=1.6$, $\omega=1$, and $q=-2a_1$.}
\end{figure}

\begin{figure}[htb]
\centering
\subfigure[]{\includegraphics[height=40mm,width=48mm]{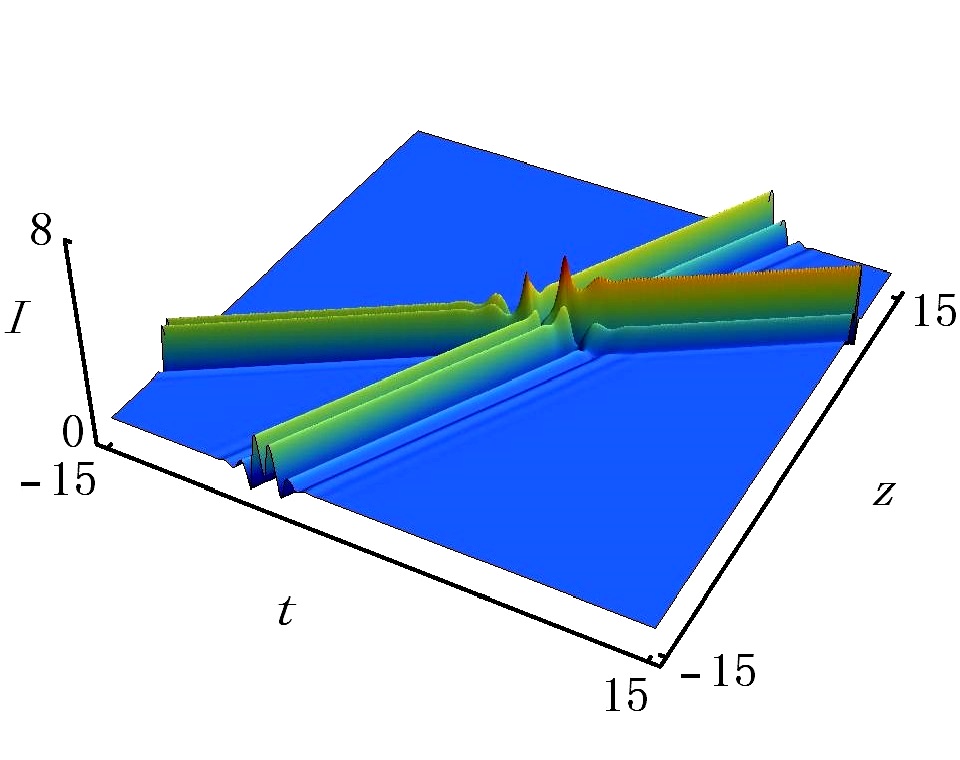}}
\subfigure[]{\includegraphics[height=40mm,width=36mm]{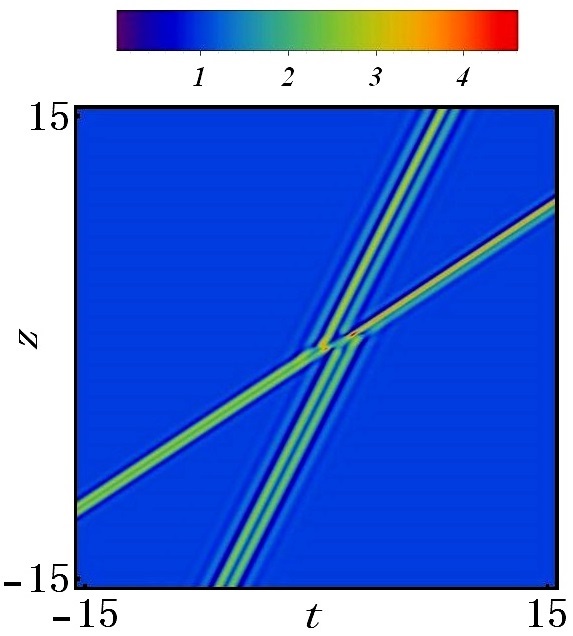}}
\subfigure[]{\includegraphics[height=35mm,width=60mm]{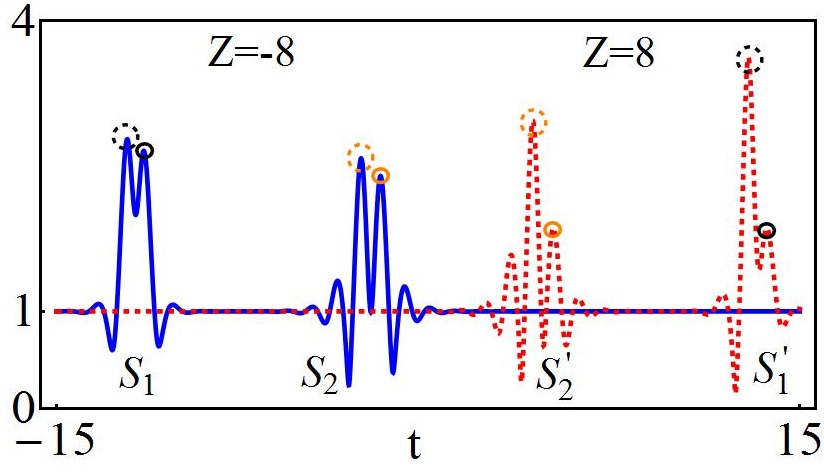}}
\caption{Shape-changing collision between two multi-peak solitons (incident solitons $S_1$, $S_2$ and outgoing solitons $S'_1$, $S'_2$)
with the conditions $b_j^2+A_j^2=k$, $q\neq-2a_j$, $j=1,2$. In (c), the dashed circles represent the main peaks with maximum intensities, and the solid
circles are the subpeaks with smaller intensities.
The setup is $a=1$, $a_1=0$, $b_1=1.5$, $a_2=0.5$, $b_2=1$, $q=5$, $\omega=1$.}
\end{figure}
\begin{table}[!hbp]
 \label{table1}
 \begin{tabular}{|c|c|c|c|}
 \hline
  Position   & Total $I_e$ & Multi-peak $S_1$ $I_e$ & Multi-peak $S_2$ $I_e$\\
  \hline
  $z=-15$ & 9.47891 & 5.66666 & 3.81225\\
  \hline
  $z=-10$ & 9.47891 & 5.66661 & 3.8123 \\
  \hline
  $z=10$ & 9.47891 &  5.66665 & 3.81225 \\
  \hline
  $z=15$ & 9.47891 & 5.66666 & 3.81225 \\
  \hline
\end{tabular}
\caption{Numerical integration verification of energy $I_e$ in Fig. (3) at different $z=-15,-10,10,15$.}
\end{table}

\subsection{Shape-changing interactions}
Let us now pay our attentions to the collision between multi-peak solitons and the interactions between multi-peak solitons
and other-type nonlinear waves. The interesting finding is that, all these interactions in which the multi-peak soliton is involved, exhibit
self-shaping characteristics. Namely, the features of the multi-peak soliton, such as the multi-peak features, and the intensity distribution,
are changed significantly before and after the interaction, while the properties of the other nonlinear waves remain invariant.

We first study the mutual collision between two multi-peak solitons with the compatibility condition of the multi-peak soliton,
i.e., $b_j^2+A_j^2=k$, $q\neq-2a_j$ ($j=1,2$).
As shown in Fig. 3, two incident multi-peak solitons $S_1$, $S_2$ with different peak distributions
[see intensity profiles in Fig. 3(c)] move from $z\rightarrow-\infty$ and
approach each other; they undergo collision around $(z,t)=(0,0)$ and form higher peaks.
They then separate after the collision and the intensity profiles
of the outgoing multi-peak structures $S'_1$, $S'_2$ distinct from the ones
of $S_1$, $S_2$. Remarkably, it is shown, the numbers of the peaks stay essentially
the same either $S_1$ ($S'_1$) or $S_2$ ($S'_2$),
but the intensities are redistributed in a striking way. Specifically, by comparison of the peak intensities
of $S_1$, $S'_1$ (or $S_2$, $S'_2$),
the intensities of main peaks (dashed circles) increase while the intensities of subpeaks (solid circles)
decrease after the collision [see Fig 3(c)]. Namely, this fascinating shape-changing collision stems from
the intensity transfer from subpeaks to main peaks of the multi-peak soliton itself. In contrast to the shape-changing collisions between standard solitons in the coupled
NLS systems that
describe a process of energy transference between solitons \cite{t1,t2,t3},
we may regard this shape-changing property as self-shaping effects of the multi-peak soliton.

\begin{figure}[htb]
\centering
\subfigure[]{\includegraphics[height=42mm,width=48mm]{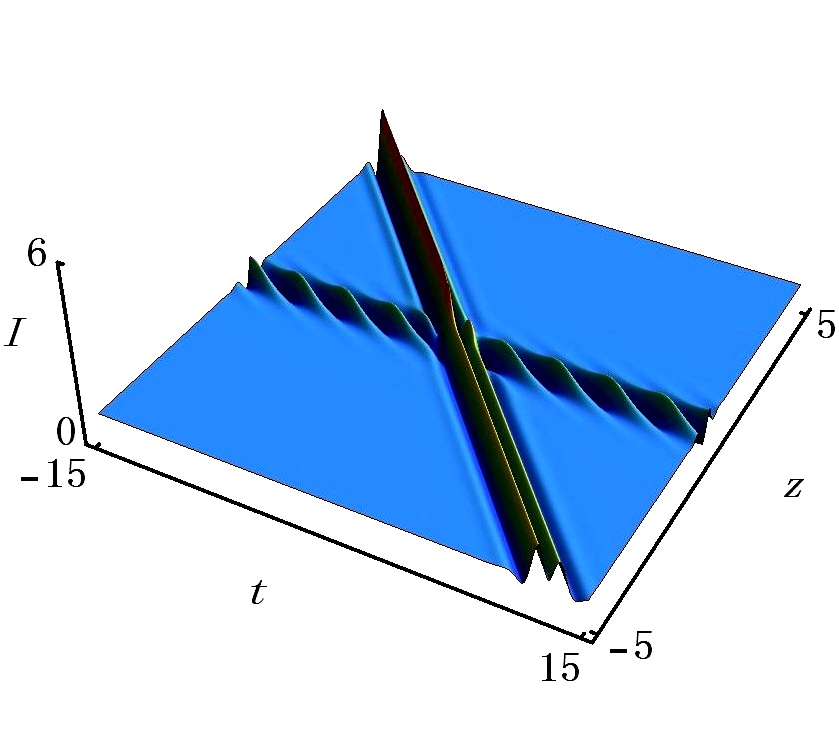}}
\subfigure[]{\includegraphics[height=42mm,width=36mm]{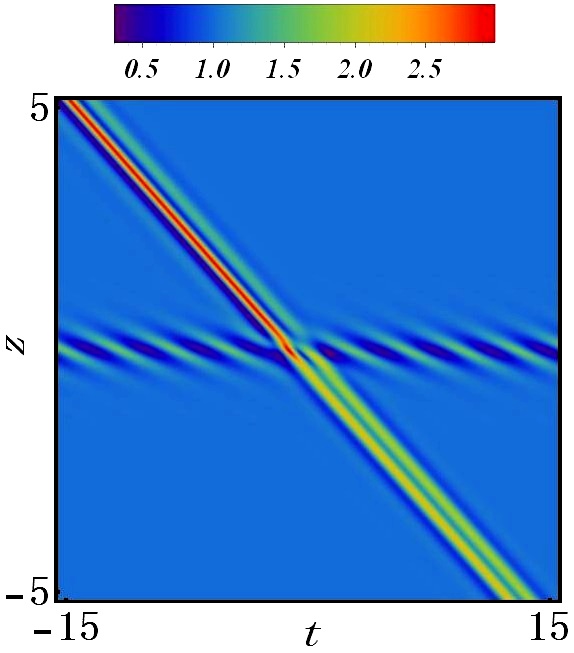}}
\subfigure[]{\includegraphics[height=35mm,width=60mm]{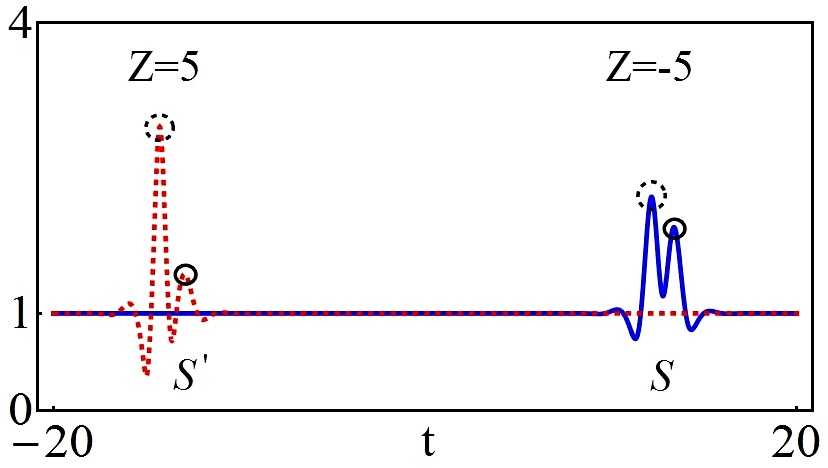}}
\caption{Shape-changing interaction between the Akhmediev breather and a multi-peak soliton (the incident soliton $S$, and outgoing soliton $S'$).
In (c), the dashed circles represent the main peaks with maximum intensities, and the solid
circles are the subpeaks with smaller intensities.
The setup is $a=1$, $a_1=-1/3$, $b_1=0.3$, $a_2=1$, $b_2=-1.1$, $q=-2a_1$, and $\omega=1$.}
\end{figure}
\begin{table}[!hbp]
 \label{table1}
 \begin{tabular}{|c|c|c|c|}
 \hline
  Position & Multi-peak soliton $I_e$\\
  \hline
  $z=-15$ & 3.75045\\
  \hline
  $z=-10$ & 3.75045\\
  \hline
  $z=10$ & 3.75045\\
  \hline
  $z=15$ & 3.75045\\
  \hline
\end{tabular}
\caption{Numerical integration verification of energy $I_e$ in Fig. (4) at different $z=-15,-10,10,15$.}
\end{table}

In order to reveal further the collision characteristics between two multi-peak solitons, we introduce the energy of light pulse against the plane-wave background with the form
\begin{equation}
I_e=\int_{t_1}^{t_2}\left\{|E(t,z)|^2-a^2\right\}dt,
\end{equation}
resulting in that the localized energy of the multi-peak solitons before and after the collision can be given clearly via optimal numerical method at different $z$.
The total energy of the two solitons is calculated numerically when $t_1=-\infty$, $t_2=\infty$. As shown in Table \uppercase \expandafter {\romannumeral 2}, the total localized energy remains invariable at $z=-15, -10, 10, 15$. The energy of the single multi-peak soliton [$I_e(s_1)$, $I_e(s_2)$] is presented by the appropriate choice for $t_1$, $t_2$.
At a selected initial propagation distance, i.e., $z<0$, we calculate $I_e(s_1)$ and $I_e(s_2)$ by $\int_{-\infty}^{t_0}\left\{|E|^2-a^2\right\}dt$ and  $\int_{t_0}^{\infty}\left\{|E|^2-a^2\right\}dt$, respectively. Note that the transverse position $t_0$ is located between $S_1$, $S_2$, leading to $|E(t_0,z)|= a$, and vice versa for positive $z$. One can see from Table \uppercase \expandafter {\romannumeral 2} that each multi-peak soliton preserves the localized energy conservation before and after the collision.
Namely, when multi-peak solitons collide, the shape change and energy conservation coexist.

For a better understanding of the interesting self-shaping effects of multi-peak solitons in the NLS-MB system, we
explore next interspecific interactions, i.e., the interactions between multi-peak solitons and other-type localized nonlinear waves.
Our aim is to demonstrate that when the interspecific interaction occurs, the multi-peak solitons exhibit self-shaping intensity distribution
but the characteristics of other-type nonlinear waves remain invariant.

\begin{figure}[htb]
\centering
\subfigure[]{\includegraphics[height=40mm,width=48mm]{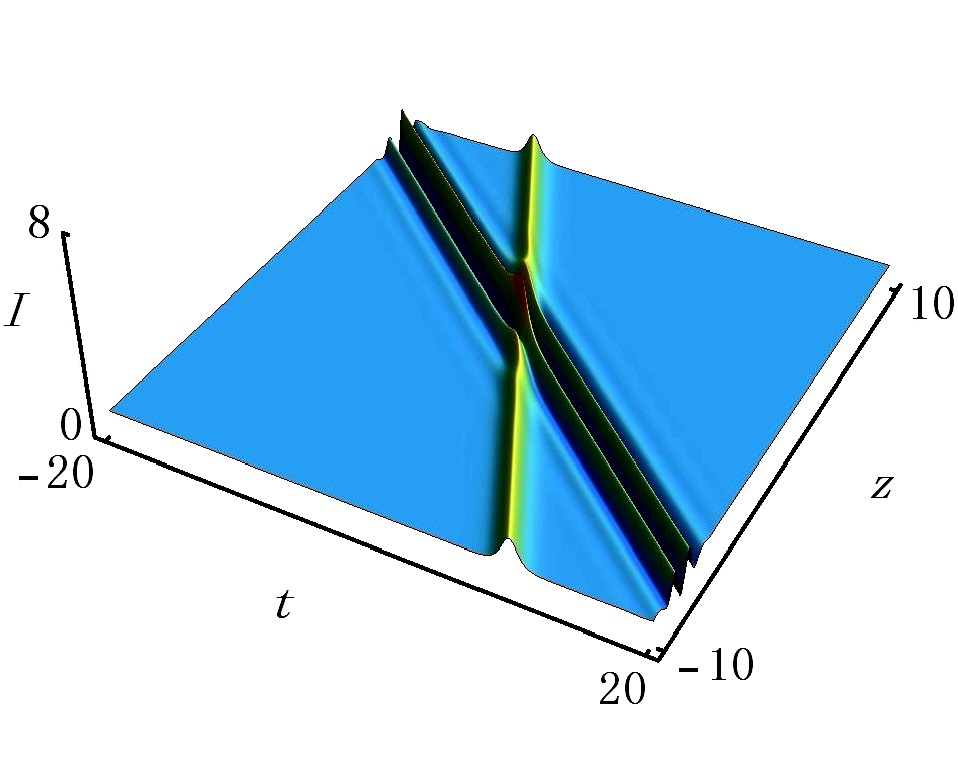}}
\subfigure[]{\includegraphics[height=40mm,width=36mm]{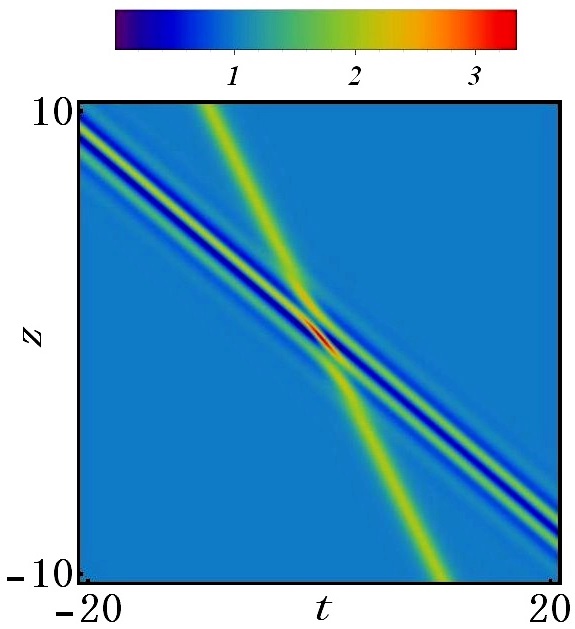}}
\subfigure[]{\includegraphics[height=35mm,width=60mm]{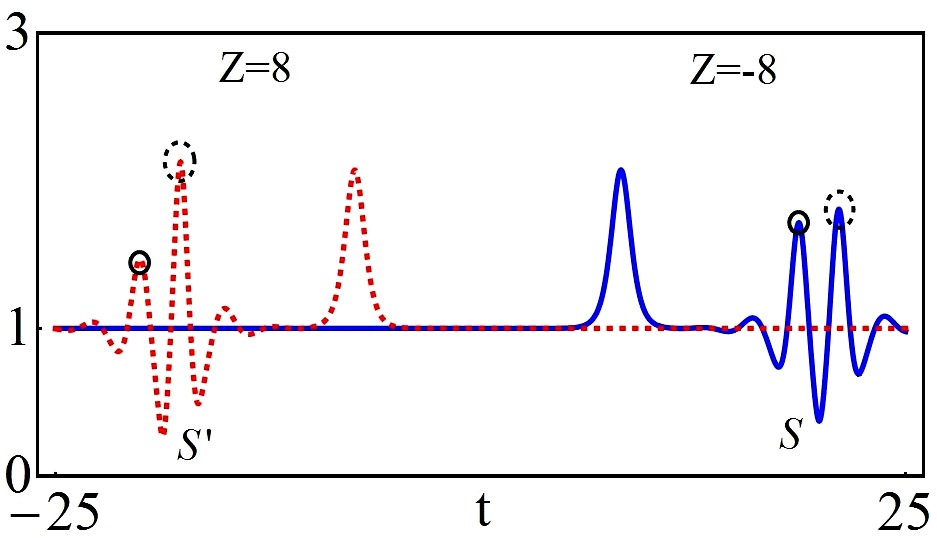}}
\caption{Shape-changing interaction between a multi-peak soliton and an anti-dark soliton.
In (c), the dashed circles represent the main peaks with maximum intensities, and the solid
circles are the subpeaks with smaller intensities. The setup is $a=1$, $a_1=0$, $b_1=1.5$, $a_2=2/3$, $b_2=\sqrt{17}/6$, $\omega=1$, and $q=-2a_1$.}
\end{figure}

\begin{table}[!hbp]
 \label{table1}
 \begin{tabular}{|c|c|c|c|}
 \hline
 Position   & Total $I_e$ & Anti-dark $I_e$ & Multi-peak $I_e$\\
  \hline
  $z=-20$ & 6.17792 & 4.47212 & 1.70580\\
  \hline
  $z=-15$ & 6.17792 & 4.47212 & 1.70580 \\
  \hline
  $z=15$ & 6.17792 & 4.47210 & 1.70582 \\
  \hline
  $z=20$ & 6.17792 & 4.47211 & 1.70582 \\
  \hline
\end{tabular}
\caption{Numerical integration verification of energy $I_e$ in Fig. (5) at different $z=-20,-15,15,20$.}
\end{table}
We first consider the interaction between multi-peak solitons and breathers for the choice of the parameters
$b_1^2+A_1^2\neq k$ and $b_2^2+A_2^2=k$, $q\neq-2a_2$. To illuminate the interaction property in a more clear way, we illustrate a
typical collision between a multi-peak soliton and a AB. As shown in Fig. 4, an incident multi-peak soliton $S$ propagating along the distance passes through the AB
near $z=0$; after that the outgoing soliton $S'$ exhibits a different intensity distribution with a higher main peak but the AB remains unchanged.
An analysis of the intensity profiles $S$ and $S'$ in Fig. 4(c) shows that, the single multi-peak soliton allows the intensity
to be transferred from the subpeak to the main peak when it collides with the AB. It is noted that a complete reverse self-shaping process occurs with the initial condition $|b_2|=-b_2$, in which a multi-peak soliton shifts the intensity from its main peak to sub peak.

Let us analyze the energy of the multi-peak soliton before and after passing through the AB by numerical integration of Eq. (2).
As shown in Table \uppercase \expandafter {\romannumeral 3}, the energy of the multi-peak soliton remains unchanged at different $z$, before and after the multi-peak soliton passes through the AB.

Another interesting case is the interplay of the multi-peak solition and antidark solitons. The corresponding initial parameter condition is extracted exactly as
$b_1^2+A_1^2=k$, $q=-2a_1$, $a^2<b_1^2$ and $b_2^2+A_2^2=k$, $q\neq-2a_2$. Figure 5 shows the structure distribution of
the collision between a multi-peak soliton and an anti-dark soliton. It is evident that two solitons are well
separated before and after the collision, which is similar to the elastic collision between the standard solitons reported before.
However, the interesting finding emerges when we look in some detail into properties of the collision.
As shown in Fig. 5(c), multi-peak soliton shows different intensity distributions before and after the collision due to its self-shaping effects,
while the anti-dark soliton remains unchanged. This indicates that there is no energy exchange between these two different-type solitons
during the collision.
Also, the corresponding numerical integration verification is displayed in Table \uppercase \expandafter {\romannumeral 4}.

It is noteworthy that, the anti-dark soliton can be regarded strictly as the limiting case of multi-peak solitons without periodic modulation
along the propagation direction, as reported before in our previous work. Thus the unique self-shaping effect is closely associated with
the periodicity of multi-peak solitons, since it does not exist in anti-dark solitons.

Furthermore,
a comparison of Figs. 3-5 implies that,
when the NLS-MB multi-peak soliton collides with another localized waves, the multi-peak pulse undergoes self-shaping phenomenon;
its intensity is redistributed in a regular way; but the other-type waves recover original shapes.
We could not detect any energy exchange between the two localized nonlinear waves. We then could infer
that the self-shaping effect is specific to the multi-peak solitons in the NLS-MB system.

\section{conclusion}
In summary, we have investigated abundant intriguing interaction structures of nonlinear waves on a plane-wave background in an erbium-doped fiber system.
In contrast to the previous results that are mainly about intraspecific interactions (solition and breather),
our results have shown that rich different types of nonlinear waves can coexist and interact with each other.
These interactions involve the shape-unchanging collisions and the shape-changing collisions.
In particular, a shape-changing collision between multi-peak solitons has been revealed as a result of the self-shaping effects of the multi-peak solitons.
It shows that the shape-changing collision preserves the power of each multi-peak soliton, which is verified via optimal numerical integration for the energy of light pulse against the plane-wave background.
Our results demonstrate that when the interactions occur, the multi-peak solitons exhibit the coexistence of the shape change and energy conservation.
It is expected that these results will enrich our understanding of the interplay of different types
of nonlinear waves in a coupled multicomponent system.

\emph{Note added}. Recently, a mathematical derivation of analytical higher-order nonlinear wave solutions of the NLS-MB system with some higher-order effects was
obtained by Wang \emph{et al} \cite{wang}.

\section*{ACKNOWLEDGEMENTS}
This work has been supported by the National Natural Science Foundation of China (NSFC) (Grant Nos. 11475135, 11547302),
and the ministry of education doctoral program funds (Grant No. 20126101110004).
C. Liu thanks T. Kanna and L.-C. Zhao for their useful discussions and valuable comments.
\section*{APPENDIX}

For the erbium-doped fiber system governed by Eq. (1), a general nonlinear wave solution in the $E$ component for arbitrary order $j$ is recursively constructed by
\begin{equation}
E_{j}=E_{j-1}-2i\frac{(\lambda_{j}-\lambda_{j}^{*})\psi_{{1},{j}}\psi_{{2},{j}}}{|\psi_{{1},{j}}|^2+|\psi_{{2},{j}}|^2},
\end{equation}
where $\psi_{{1},{j}}$ and $\psi_{{2},{j}}$ are eigenfunctions corresponding to $j$-fold Darboux transformation. In order to obtain rich nonlinear wave solutions on a background of Eq. (1), we first introduce the following plane-wave background with a generalized form
\begin{eqnarray}
&&E_0=ae^{i\theta},~P_0=i k E_0,~\eta_0=\omega k-q k/2,\nonumber
\end{eqnarray}
where $\theta=q t+\nu z$, $\nu=a^2+2k-q^2/2$,
$a$ and $q$ represent the amplitude and frequency of background electric field, respectively,
and $k$ is a real parameter which is related to the background amplitude of $P$ component. By solving the Lax pair in the Ref. \cite{p3}, the eigenfunctions $\psi={\psi_{1,1}\choose \psi_{2,1}}$ corresponding to one-fold Darboux transformation with the eigenvalues $\lambda_{1}=a_1+ib_1$ is given
\begin{eqnarray}
&&\psi_{1,1}(\lambda_{1})=a(\phi_{1}+\phi_{2}),\nonumber\\
&&\psi_{2,1}(\lambda_{1})=[i\lambda_{1}(\phi_{1}+\phi_{1})+(\tau_{1}\phi_{1}+\tau_{2}\phi_{2})],~~
\end{eqnarray}
where
\begin{eqnarray}
&&\phi_{1}=\exp[\tau_{1}t+(i\tau_{1}^2+B+C)z],\nonumber\\
&&\phi_{2}=\exp[\tau_{2}t+(i\tau_{2}^2+B+C)z],\nonumber\\
&&\tau_{1}=\left(iq+\sqrt{-q^2-4a^2-4\lambda_{1} q-4\lambda_{1}^2}\right)/2,\nonumber\\
&&\tau_{2}=\left(iq-\sqrt{-q^2-4a^2-4\lambda_{1} q-4\lambda_{1}^2}\right)/2,\nonumber\\
&&B=q/2+\lambda_{1}+k/(\lambda_{1}+\omega),\nonumber\\
&&C=\lambda_{1}^2+\lambda_{1} q/2+3a^2/2-kq/(2\lambda_{1}+2\omega),
\end{eqnarray}
Substitute the eigenfunctions $\psi_{1,1}(\lambda_1)$ and $\psi_{2,1}(\lambda_1)$ into the Eq. (A1) with a tedious simplification, the general and concise expression of the first-order nonlinear wave solution is presented as
\begin{equation}
E_{1}=E_{0}\left\{1-\frac{8b_1m_{1}[\sin{(\gamma+\mu_1)}-i\sinh{(\beta+i\mu_1)}]}
 {m_3\sin{(\gamma+\mu_2)}-im_2\sinh{(\beta-i\mu_3)}}\right\},
\end{equation}
where
\begin{eqnarray}
&&\beta=\zeta(t+V_1 z),~\gamma=\sigma(t+V_2 z),\nonumber\\
&&V_1=v_1+b_1 \sigma v_2/\zeta,~V_2=v_1-b_1 \zeta v_2/\sigma,\nonumber\\
&&v_1=k (a_1+\omega)/[b_1^2+(a_1+\omega)^2]+(a_1-q/2),\nonumber\\
&&v_2=1-k/[b_1^2+(a_1+\omega)^2],\nonumber\\
&&\zeta=\left(\sqrt{\chi^2+\mu^2}+\chi\right)^{1/2}/\sqrt{2},\nonumber\\
&&\sigma=\pm\left(\sqrt{\chi^2+\mu^2}-\chi\right)^{1/2}/\sqrt{2},\nonumber\\
&&\chi=4b_1^2-4a^2-(2a_1+q)^2,~\mu=-4b_1(2a_1+q),\nonumber\\
&&m_1=\left\{(i\zeta-\sigma)^2+[2b_1+i(2a_1+q)]^2\right\}^{1/2},\nonumber\\
&&m_2=\left\{(\alpha_1+\alpha_2)^2-4[2b_1\zeta+(2a_1+q)\sigma]^2\right\}^{1/2},\nonumber\\
&&m_3=\left\{(\alpha_1-\alpha_2)^2+4[2b_1\sigma-(2a_1+q)\zeta]^2\right\}^{1/2},\nonumber\\
&&\alpha_1=4a^2+4b_1^2+(2a_1+q)^2,~\alpha_2=\zeta^2+\sigma^2,\nonumber\\
&&\tan{\mu_1}=\frac{2b_1+i(2a_1+q)}{\sigma-i\zeta},\nonumber\\
&&\tan{\mu_2}=\frac{\alpha_1-\alpha_2}{4b_1\sigma-2(2a_1+q)\zeta},\nonumber\\
&&\tan{\mu_3}=\frac{\alpha_1+\alpha_2}{4b_1\zeta+2(2a_1+q)\sigma}.
\end{eqnarray}
Here $`\pm'$ in $\sigma$ depends on $\mu\leq0$ and $\mu>0$, respectively.
The higher-order nonlinear wave solution is generated by the iteration of Darboux transformation. Here we present the second-order solution
with the eigenfunctions $\psi_{1,2}$ and $\psi_{2,2}$ of two-fold Darboux transformation which are given
\begin{eqnarray}
&&\psi_{{1},{2}}=(\lambda_2-\lambda_1^*-2ib_1M)\psi_{{1},{1}}(\lambda_2)-2ib_1N\psi_{2,1}(\lambda_2),\nonumber\\
&&\psi_{{2},{2}}=(\lambda_2-\lambda_1+2ib_1M)\psi_{2,1}(\lambda_2)-2ib_1N^*\psi_{1,1}(\lambda_2).\nonumber
\end{eqnarray}
where
\begin{eqnarray}
&&M=\frac{4a^2(\cosh{\beta}+\cos{\gamma})}
 {m_3\sin{(\gamma+\mu_2)}-im_2\sinh{(\beta-i\mu_3)}},\nonumber\\
&&N=\frac{2am_{1}[\sin{(\gamma+\mu_1)}-i\sinh{(\beta+i\mu_1)}]}
 {m_3\sin{(\gamma+\mu_2)}-im_2\sinh{(\beta-i\mu_3)}}e^{i\theta}.
\end{eqnarray}


\begin{thebibliography}{99}
\bibitem{NLW1} J. Yang, \emph{Nonlinear Waves in Integrable and Nonintegrable Systems} (SIAM,
Philadelphia, 2010).
\bibitem{NLW2} N. Akhmediev and A. Ankiewicz, \emph{Solitons: Nolinear Pulses and Beams} (Chapman and Hall, London, 1997).
\bibitem{NLW3} P. G. Kevrekidis, D. Frantzeskakis, and R. Carretero-Gonzalez,
\emph{Emergent Nonlinear Phenomena in Bose-Einstein Condensates: Theory and Experiment} (Springer, Berlin Heidelberg, 2009).
\bibitem{NLW4} G.P. Agrawal, \emph{Nonlinear Fiber Optics} (4th Edition, Acdemic Press, Boston, 2007).
\bibitem{s1} G. I. Stegeman and M. Segev, Science, 286, 1518 (1999).
\bibitem{s2} W. J. Liu, B. Tian, H. Q. Zhang, L. L. Li, and Y. S. Xue, Phys. Rev. E \textbf{77}, 066605 (2008).
\bibitem{s3} A. Chowdury, D. J. Kedziora, A. Ankiewicz, and N. Akhmediev, Phys. Rev. E 90, 032922 (2014).
\bibitem{b1} N. Akhmediev, J. M. Soto-Crespo, and A. Ankiewicz, Phys. Rev. A \textbf{80}, 043818 (2009).
\bibitem{b2} D. J. Kedziora, A. Ankiewicz, and N. Akhmediev, Phys. Rev. E \textbf{85}, 066601 (2012).
\bibitem{b3} J. S. He, H. R. Zhang, L. H. Wang, K. Porsezian, and A. S. Fokas, Phys. Rev. E \textbf{87}, 052914 (2013).
\bibitem{b4} V. E. Zakharov and A. A. Gelash, Phys. Rev. Lett. 111, 054101 (2013).
\bibitem{b5} A. Chowdury, D. J. Kedziora, A. Ankiewicz, and N. Akhmediev, Phys. Rev. E 91, 022919 (2015).
\bibitem{b6} B. Frisquet, B. Kibler, and G. Millot, Phys. Rev. X \textbf{3}, 041032 (2013).
\bibitem{b7} B. Kibler, A. Chabchoub, A. Gelash, N. Akhmediev, and V. E. Zakharov, Phys. Rev. X 5, 041026 (2015).
\bibitem{r1} C. Becker, S. Stellmer, P. S. Panahi, S. Dorscher,
M. Baumert, Eva-Maria Richter, J. Kronjager, K. Bongs, and K. Sengstock, Nat. Phys. \textbf{4}, 496-501 (2008).
\bibitem{r2} K.W. Chow, R.H.J. Grimshaw, and E. Ding, Wave Motion 43, 158¨C166 (2005).
\bibitem{r3} T. Xu, D. Wang, M. Li, and H. Liang, Phys. Scr. 89, 075207 (2014).
\bibitem{r4} A. Chowdury, D. J. Kedziora, A. Ankiewicz, and N. Akhmediev, Phys. Rev. E 91, 032928 (2015).
\bibitem{r5} H. J. Shin, Phys. Rev. E 63, 026606 (2001); 71, 036628 (2005); J. Phys. A: Math. Theor. 45, 255206 (2012).
\bibitem{r6} X. P. Cheng, S. Y. Lou, C. L. Chen, and X. Y. Tang, Phys. Rev. E  89, 043202 (2014).
\bibitem{r7} D. J. Kedziora, A. Ankiewicz, N. Akhmediev, Eur. Phys. J. Spec. Top. 223, 43 (2014).
\bibitem{t1} T. Kanna and M. Lakshmanan, Phys. Rev. Lett. \textbf{86}, 5043 (2001); Phys. Rev. E \textbf{67}, 046617 (2003);
M. Vijayajayanthi, T. Kanna, and M. Lakshmanan, Phys. Rev. A \textbf{77}, 013820 (2008); S. Rajendran, M. Lakshmanan,
and P. Muruganandam, J. Math. Phys. 52, 023515 (2011); T. Kanna, K. Sakkaravarthi, and M. Vijayajayanthi, Pramana J. Physics, \textbf{85}, 881 (2015).
\bibitem{t2} Y. Jiang, B. Tian, W. J. Liu, K. Sun, M. Li, and P. Wang, Phys. Rev. E \textbf{85}, 036605 (2012).
\bibitem{t3} D. S. Wang, D. J. Zhang, J. Yang, J. Math. Phys. 51, 023510 (2010).
\bibitem{n1} C. Kalla, J. Phys. A 44 (2011) 335210.
\bibitem{n2} B. L. Guo and L. M. Ling, Chin. Phys. Lett. 28 (2011) 110202.
\bibitem{n3} F. Baronio, A. Degasperis, M. Conforti, and S. Wabnitz, Phys. Rev. Lett. 109 (2012) 044102.
\bibitem{n4} L. C. Zhao and J. Liu, J. Opt. Soc. Am. B 29, 3119 (2012).
\bibitem{n5} A. Degasperis and S. Lombardo, Phys. Rev. E 88, 052914 (2013).
\bibitem{n6} L. M. Ling, L. C. Zhao, B. L. Guo, arXiv:1407.5194 [nlin.SI] (2014).
\bibitem{n7} C. Liu, Z. Y. Yang, L. C. Zhao, and W. L. Yang, Phys. Rev. A 89, 055803 (2014).
\bibitem{n8} C. Liu, Z. Y. Yang, L. C. Zhao, and W. L. Yang, Ann. Phys. 362, 130 (2015).
\bibitem{n9} L. C. Zhao, Z. Y. Yang, and L. M. Ling, J. Phys. Soc. Jpn. 83, 104401 (2014).
\bibitem{n10} L. M. Ling and L. C. Zhao, Phys. Rev. E 92, 022924 (2015).
\bibitem{m1} A. I. Maimistov and A. M. Basharov, Nonlinear Optical Waves
(Springer-Verlag, Berlin, 1999).
\bibitem{m2} S. Kakei and J. Satsuma, J. Phys. Soc. Jpn. 63 (1994) 885.
\bibitem{m3} K. Porsezian and K. Nakkeeran, J. Mod. Opt. 42 (1995) 1953;
K. Porsezian and K. Nakkeeran, Phys. Rev. Lett. 74 (1995) 2941;
K. Porsezian J. Mod. Opt. 47 (2000) 1635.
\bibitem{p1} J. S. He, S. W. Xu, and K. Porsezian, Phys. Rev. E 86 (2012) 066603;
C. Z. Li, J. S. He, K. Porsezian, Phys. Rev. E 87 (2013) 012913.
\bibitem{p2} R. Guo, B. Tian, X. L\"{u}, H. Q. Zhang, and W. J. Liu Comput. Math. Phys. 52 (2012) 565.
\bibitem{p3} R. Guo, H. Q. Hao and L. L. Zhang, Mod. Phys. Lett. B 27 (2013) 1350130.
\bibitem{p4} R. Guo and H. Q. Hao, Commun. Nonlinear Sci.
Numer. Simulat. 19 (2014) 3529; Ann. Phys. 344 (2014) 10.
\bibitem{p5} L. Wang, X. Li, F. H. Qi, and L. L. Zhang, Ann. Phys. 359, 97 (2015).
\bibitem{chong} Y. Ren, Z. Y. Yang, C. Liu, and W. L. Yang, Phys. Lett. A 379, 2991 (2015) .
\bibitem{AB} N. Akhmediev and V. I. Korneev, Theor. Math. Phys. 69 (1986) 1089-1093.
\bibitem{KMB} E. Kuznetsov, Sov. Phys. Dokl. 22 (1977) 507; Y. C. Ma, Stud. Appl. Math. 60 (1979) 43-58.
\bibitem{wang} L. Wang, Y. J. Zhu, J. H. Zhang, T. Xu, F. H. Qi, and Y. S. Xue, J. Phys. Soc. Jpn. \textbf{85}, 024001 (2016).
\end{thebibliography}
\end{document}